\documentclass[conference]{IEEEtran}

\usepackage{cite}
\usepackage{url}
\usepackage{amsmath,amssymb}
\usepackage{graphicx}
\usepackage{booktabs}
\usepackage{multirow}
\usepackage{pgfplots}
\pgfplotsset{compat=1.18}
\usepackage{tikz}
\usetikzlibrary{calc,arrows.meta,positioning}
\usepackage{float}
\usepackage{subcaption}
\usepackage{tabularx}

\begin{document}

\title{Predictive Sectorization and Bayesian Optimized\\
Consensus for Admission Control in\\Autonomous Airspace Operations}

\author{
\IEEEauthorblockN{Aditya Dhodapkar\IEEEauthorrefmark{1},
Avery Smidt\IEEEauthorrefmark{2},
Aaron Verkleeren\IEEEauthorrefmark{1},
Stacy Patterson\IEEEauthorrefmark{1},
Carlos A.\ Varela\IEEEauthorrefmark{1}}
\IEEEauthorblockA{\IEEEauthorrefmark{1}Department of Computer Science,
Rensselaer Polytechnic Institute, Troy, NY, USA\\
\IEEEauthorrefmark{2}Quaternion Consulting Inc., Herndon, VA, USA\\
\{dhodaa, verkla\}@rpi.edu, asmidt@quaternion-consulting.com, \{sep, cvarela\}@cs.rpi.edu}
}

\maketitle

\begin{abstract}
Conventional air traffic control divides airspace into specific
regions, creating a scaling bottleneck as traffic grows.  Choosing how to partition airspace
is not straightforward because grid size affects workload, handoff
frequency, and the capacity of whatever coordination mechanism operates
within each sector.  We present a three stage pipeline that automates
sectorization and sector coordination while preserving human oversight.  First, a two
stage XGBoost classifier predicts the optimal 3D grid configuration from
23 location-agnostic traffic features, achieving 91.38\% accuracy on a
65,000 sample dataset derived from Federal Aviation Administration
System Wide Information Management replays.
Second, a leaderless Paxos consensus protocol lets aircraft coordinate
sector entries among themselves, maintaining above 96\% entry success
with low near mid-air collision rates across all tested configurations.
Third, Bayesian Optimization with a Gaussian Process
surrogate tunes eight protocol parameters per airport in 50 trials,
revealing that each
traffic environment requires a qualitatively different configuration.
The resulting pipeline offers a practical path toward scalable,
autonomous airspace management as traffic demand outpaces controller
capacity.
\end{abstract}

\section{Introduction}

Air traffic control operates reliably almost all the time.  The problem
is what happens when it does not.  The Federal Aviation Administration (FAA) has reported staffing
shortages at major facilities for
years~\cite{dot_oig_av2023035_2023, faa_cwp_2025}.  Controllers work
overtime and six day weeks to maintain coverage.  In Europe, industrial
action has repeatedly produced network level delays and
cancellations~\cite{eurocontrol_nor_2023}.  Traffic surges and weather induced rerouting compound controller workload
simultaneously, increasing handoff frequency, reducing separation
margins, and shortening available decision time.

In the air traffic control system, one controller is responsible for every
aircraft in their sector.  A controller typically manages 8--12 aircraft
simultaneously, a manageable cognitive load, but peak periods can push
this above 18, producing an overwhelming workload.
Research on dynamic density~\cite{nasa_tm_1998_dynamic_density} and
cognitive complexity~\cite{eurocontrol_cogcomplex_2003} has shown that
workload depends not just on aircraft count but on proximity, closure
rates, heading changes, and crossing streams.  Dynamic Airspace
Configuration (DAC) has been proposed as a
remedy~\cite{kopardekar_dac_2007}, adjusting sector boundaries in
response to evolving demand.  But existing DAC approaches rely on
computationally expensive graph cut or clustering
optimizations~\cite{polishchuk2011_survey} that are difficult to run at
the cadence required for tactical decisions.

Sectorization research spans algorithmic boundary
placement~\cite{polishchuk2011_survey} and demand responsive
configuration~\cite{kopardekar_dac_2007}, both relying on workload
metrics rooted in dynamic density~\cite{nasa_tm_1998_dynamic_density}
and cognitive complexity~\cite{eurocontrol_cogcomplex_2003}.  For
decentralized coordination, Paul et al.\ introduced conflict aware
flight planning~\cite{paul_dasc_2019, paul_dasc_2020} with
formal verification of safety under asynchronous
networks~\cite{paul_isse_2022, paul_scp_2025}.  We build on this work
in our Decentralized Air Traffic Control (DATC) protocol.  Bayesian Optimization
has been applied to expensive black box tuning across
domains~\cite{shahriari_bo_survey_2016, snoek_practical_bo_2012}, but
not to consensus protocol parameters in aviation.

We treat sectorization as a supervised classification problem and pair
it with a decentralized consensus protocol, keeping controllers in
charge of policy while offloading routine separation tasks.

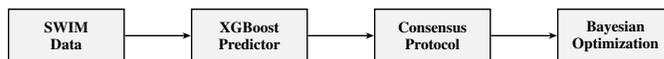
\begin{figure}[!htb]
\centering
\resizebox{\columnwidth}{!}{%
\begin{tikzpicture}[
  box/.style={rectangle, draw=black, thick, fill=gray!10,
              minimum width=2.4cm, minimum height=1.1cm, align=center,
              font=\bfseries\small},
  arr/.style={-{Stealth[length=5pt]}, thick}
]
\node[box] (sim)  at (0,0)    {SWIM\\Data};
\node[box] (pred) at (3.8,0)  {XGBoost\\Predictor};
\node[box] (cons) at (7.6,0)  {Consensus\\Protocol};
\node[box] (tune) at (11.4,0) {Bayesian\\Optimization};
\draw[arr] (sim)  -- (pred);
\draw[arr] (pred) -- (cons);
\draw[arr] (cons) -- (tune);
\end{tikzpicture}%
}
\caption{Pipeline overview.}\label{fig:pipeline}
\end{figure}

Our proposed pipeline (Fig.~\ref{fig:pipeline}) has three stages.
\textbf{Stage~1:} A two stage XGBoost predictor takes as input 23
aggregate traffic features derived from a snapshot of the airspace
(density, proximity, flow direction, altitude mix, etc.) and maps them
to the optimal grid configuration, i.e., the number of rows and columns
of a uniform rectangular partition that divides the airspace into
sectors.
\textbf{Stage~2:} The predicted grid defines sector boundaries for a
leaderless Paxos consensus protocol in which aircraft already occupying
a sector coordinate with an arriving aircraft to decide whether its
proposed flight plan can be admitted without causing conflicts.
\textbf{Stage~3:} Bayesian Optimization (BO) with a Gaussian Process
surrogate tunes eight protocol parameters per airport to maximize
admission success while minimizing holding patterns, speed
modifications, and retries.
To our knowledge, this is the first work to combine learned sectorization,
decentralized consensus, and automatic protocol tuning into a single end
to end pipeline.

Our contributions are:
\begin{enumerate}
  \item A two stage XGBoost classifier achieving 91.38\% accuracy on 25
        class sectorization prediction using 23 location-agnostic
        features, outperforming Random Forest, LightGBM, CatBoost, and
        single stage XGBoost baselines on the same dataset.
  \item A DATC consensus protocol contribution:
    \begin{enumerate}
      \item A standalone integration of three protocols (Discovery,
            Synod, TAP)~\cite{paul_dasc_2020, paul_maes_2023} into
            the DATC consensus framework with reliability extensions.
      \item Validation on both synthetic and real JFK System Wide
            Information Management (SWIM) traffic across grid sizes
            from $2{\times}2$ to $16{\times}16$ and up to 160 aircraft.
    \end{enumerate}
  \item A Bayesian Optimization framework that tunes eight protocol
        parameters per airport in 50 trials, demonstrating that no single
        default configuration is sufficient across traffic environments.
\end{enumerate}

A full treatment of each stage, including additional baselines,
ablations, and extended evaluation, is available in the first
author's master's thesis~\cite{dhodapkar_thesis_2026}.

The remainder of this paper is organized as follows.  Section~II
presents the sectorization predictor, Section~III describes the
consensus protocol, and Section~IV covers the Bayesian Optimization
framework.  Section~V discusses findings and limitations, and
Section~VI concludes.

\section{Sectorization Predictor}

\subsection{Problem and Data}

Given a traffic snapshot over a $100 \times 100$\,Nautical Mile (NM)
study region, we
predict the optimal horizontal grid $(R^*,C^*)$ from $\{1,\dots,5\}^2$
(25 configurations) with three fixed altitude layers (0--9,999\,ft,
10,000--17,999\,ft, 18,000+\,ft).  The 100\,NM side length ensures that
even the finest predictor grid ($5{\times}5$) produces 20\,NM sectors,
large enough for adequate dwell time at cruise speeds.  Labels are generated by
exhaustive grid search over a composite objective combining occupancy
variance, handoff counts, and risk pair penalties:
\begin{equation}
(R^*,C^*) = \arg\min_{(R,C)} J(R,C,\, L{=}3).
\end{equation}

Traffic was recorded from $100 \times 100$\,NM regions centered on five
major US airports (JFK, ATL, ORD, DFW, LAX) and five en route cruising
corridors with consistently high traffic flow, via the FAA SWIM
program~\cite{faa_swim_overview}.  In most raw snapshots, traffic density
is low enough that $1{\times}1$ (no sectorization) dominates the
unbalanced dataset.  We augment underrepresented configurations with
the Synthetic Minority Over-sampling Technique
(SMOTE)~\cite{chawla_smote_2002}, producing approximately 65,000 samples
(40,000 simulation, 25,000 synthetic) split 70/15/15 for training,
validation, and testing.  Most configurations receive approximately
2,700 samples each, with $1{\times}1$ at 5,350 and $5{\times}5$ at 300.

\subsection{Features}

We extract 23 location-agnostic features organized into three groups
(Table~\ref{tab:features}).  No geographic identifiers are included,
forcing the model to learn from traffic characteristics alone.

\begin{table}[!htb]
\centering
\caption{Feature summary (23 features).}\label{tab:features}
\footnotesize
\renewcommand{\arraystretch}{0.9}
\begin{tabular}{p{2.6cm}p{3.2cm}p{1.0cm}}
\toprule
\textbf{Feature} & \textbf{Description} & \textbf{Range} \\
\midrule
\multicolumn{3}{l}{\textit{Raw (8)}} \\
traffic\_density  & Aircraft count & $[0,150]$ \\
avg\_proximity    & Mean pairwise dist. & $[0,80]$\,NM \\
altitude\_mix     & Altitude std.\ dev. & $[0,15k]$\,ft \\
conflict\_risk    & Mean min.\ sep. & $[0,60]$\,NM \\
primary\_flow\_dir & Mean $|\sin\psi|$ & $[0,1]$ \\
flow\_concentration & Circ.\ std.\ dev. & $[0,1.5]$ \\
time\_of\_day     & Hour & $[0,23]$ \\
airspace\_size    & Region radius & 100 NM \\
\midrule
\multicolumn{3}{l}{\textit{Engineered (9)}} \\
congestion\_index & density/proximity & -- \\
traffic\_alt\_complexity & density $\times$ alt\_mix & -- \\
hotspot\_indicator & risk/proximity & -- \\
density\_sq       & density$^2$ & -- \\
log\_proximity    & $\ln(1{+}\text{prox})$ & -- \\
traffic\_level    & Binned density & $\{0,1,2\}$ \\
proximity\_level  & Binned proximity & $\{0,1,2\}$ \\
risk\_normalized  & risk/density & -- \\
flow\_direction   & Axial alignment & $[0,1]$ \\
\midrule
\multicolumn{3}{l}{\textit{Time/day (6)}} \\
\multicolumn{3}{l}{day\_weekday, day\_weekend, time\_\{morn, noon, eve, night\}} \\
\bottomrule
\end{tabular}
\end{table}

\subsection{Model Architecture}

The model uses a two stage XGBoost~\cite{chen_xgboost_2016}
architecture.  Stage~1 is a binary classifier (500 trees, depth 6,
learning rate 0.05) that separates $1{\times}1$ from non-$1{\times}1$
samples with 99.29\% recall.  This near perfect recall ensures the model
almost never predicts single sector when multiple sectors are needed,
which would cause controller overload.  Stage~2 is a 24 class classifier
(800 trees, depth 14, learning rate 0.02, $L_1 = 0.1$) that predicts the
exact configuration.  At inference, a sample classified as $1{\times}1$
by Stage~1 receives $(R,C)=(1,1)$ directly; otherwise Stage~2 predicts
the full configuration.

Because $1{\times}1$ dominates raw SWIM traffic, a single stage 25 class
model learns to default to the trivial case and achieves only 85.12\%
accuracy.
The two stage design filters these out first so Stage~2 only
discriminates among configurations that actually need partitioning.
\subsection{Results}

The two stage model achieves 91.38\% overall accuracy ($91.1 \pm 0.3\%$
under five fold cross validation).  Stage~2 predicts rows correctly
95.5\% of the time and columns 94.8\%, but both must be correct
simultaneously, producing the 90.91\% combined non-$1{\times}1$
accuracy.  Table~\ref{tab:baselines} compares against four alternatives
trained on the same data with the same split.

\begin{table}[!htb]
\centering
\caption{Model comparison on the held out test set.}\label{tab:baselines}
\renewcommand{\arraystretch}{1.2}
\begin{tabular}{lr}
\toprule
\textbf{Model} & \textbf{Accuracy} \\
\midrule
Single Stage XGBoost  & 85.12\% \\
Random Forest         & 90.17\% \\
LightGBM              & 90.69\% \\
CatBoost              & 90.83\% \\
\textbf{Two Stage XGBoost} & \textbf{91.38\%} \\
\bottomrule
\end{tabular}
\end{table}

Confusion matrix analysis reveals three systematic error patterns:
(1) low partition confusion ($1{\times}2 \leftrightarrow 2{\times}1$,
both represent a single bisection differing only in orientation),
(2) row count adjacency (e.g., $5{\times}1 \to 4{\times}1$ when
density profiles overlap), and (3) column swap ($3{\times}4
\leftrightarrow 3{\times}5$, correct row with off by one column).  In
practice, these neighboring configurations produce similar workload
distributions, limiting the operational impact of such errors.  The
full 25-class confusion matrix is available in the thesis.

The top three features (traffic level, traffic density, traffic density
squared) account for over 79\% of total gain, confirming that
sectorization is primarily driven by traffic volume.  Inference requires
less than 1\,ms per prediction on a single CPU core.

\section{Consensus Protocol}

\subsection{Background and Architecture}

The predicted grid defines sector boundaries.  Within each sector,
aircraft coordinate entries using the DATC
protocol~\cite{paul_dasc_2020, paul_maes_2023}, which adapts
Lamport's Paxos~\cite{lamport_paxos_1998} into a three phase admission
procedure.  Paul et al.\ developed conflict aware flight planning with
formally verified correctness
guarantees~\cite{paul_dasc_2019, paul_scp_2025} and proved eventual and timely consensus under specific network constraints regarding message delays, reordering, and
loss~\cite{paul_isse_2022}.

We integrated the three constituent protocols (Discovery, Synod,
TAP) into a standalone C++ library driven by a discrete event simulator
with a per aircraft engine architecture.  Each aircraft runs its own consensus
engine containing proposal state, message history, and deduplication
sets.  The engines communicate through a shared event queue ordered by
(time, aircraft ID, event type) tuples with integer picosecond
timestamps for deterministic execution.  Reliability extensions include
fuel tracking with diversion logic and randomized exponential backoff
for livelock prevention; full implementation details and the
correctness argument appear in the thesis.

\subsection{Three Phase Protocol}

Every sector admission runs through three sequential phases.

\textbf{Phase~A (Discovery):} The entering aircraft broadcasts
INIT\_REQ to discover current sector occupants.  REQ\_ACK responses
establish the quorum (majority of occupants needed to agree, ensuring
safety under message loss) and provide each occupant's current flight plan.
The proposer uses this information to compute a provably collision-free
flight plan before entering Synod.

\textbf{Phase~B (Synod):} A single round of the Synod
protocol~\cite{lamport_paxos_1998} determines whether the proposed
plan can be admitted.  The proposer sends PREPARE with a unique
proposal number; occupants respond with PROMISE, including the
number and value of the highest-numbered proposal they have already
accepted (if any).  Once a quorum promises, the proposer sends
ACCEPT: if any PROMISE carried an accepted value, the proposer must
re-propose the value with the highest proposal number; otherwise it
proposes its own collision-free plan with a certificate of
collision-freedom.  If no conflict-free plan can be constructed, a
formally verified backtracking algorithm~\cite{paul_dasc_2019}
adjusts ground speeds.  If speed adjustment fails, a holding pattern
is assigned.  If all alternatives are exhausted, the entry is denied.
When multiple aircraft propose simultaneously, they can repeatedly
preempt each other by incrementing proposal numbers in an unbroken
cycle; we mitigate this livelock risk with randomized exponential
backoff drawn from $[\texttt{nackBackoffMin},\, \texttt{nackBackoffMax}]$.

\textbf{Phase~C (TAP, Two Phase Acknowledge Protocol):} The admitted
plan is reliably disseminated to all sector members in two rounds:
LEARN/LEARNT confirms every member received the decided value, then
AK/ACK confirms every member knows that every other member received
it, ensuring every aircraft holds an identical view of the admitted
plan set before the next admission.

\textbf{Safety and liveness.}
Safety here means that no two conflicting flight plans are ever both
admitted, even under arbitrary message delays, reordering, and loss.
This is guaranteed by Paxos quorum intersection: any two majority
quorums share at least one member, so at least one voter in any new
round has seen the latest admitted
plan~\cite{lamport_paxos_1998, paul_isse_2022}.  TAP ensures all sector
members share an identical view of the admitted set, preventing
divergent views.  Our primary results address liveness: eventual
consensus progress via livelock prevention~\cite{paul_dasc_2019} and
timely TAP completion under probabilistic message
latency~\cite{paul_dasc_2020}.

\textbf{Exit protocols.}
When an aircraft leaves a sector, it runs the second two phases
(Synod and TAP) with an exit flag, proposing removal of its plan from
the admitted set rather than addition.  Discovery is unnecessary because
the exiting aircraft is already a sector member.  The Synod round agrees
on the removal, and TAP disseminates the updated state.  A denied entry aircraft that has
exhausted all alternates initiates an emergency exit (C3 exit,
abandoning its expected arrival slot) to abort
the attempt, remove itself from the sector's tracking sets, and notify
all occupants that the entry has been abandoned.  This symmetric design
means entry and exit use identical protocol machinery, simplifying both
verification and implementation.

\subsection{Evaluation}

We evaluate the protocol through parametric sweeps across grid
configurations and traffic counts on both synthetic trajectories (grid
sizes $2{\times}2$, $4{\times}4$, $8{\times}8$, $16{\times}16$; traffic
counts 10 to 160) and real JFK SWIM data (grid sizes $2{\times}2$
through $6{\times}6$; traffic counts 10 to 80).  Note that these sweep
ranges are broader than the predictor's $\{1,\dots,5\}^2$ space; the
larger grids stress test the protocol beyond predicted configurations.

All synthetic configurations maintain above 96\% entry success through
160 aircraft, with the $8{\times}8$ grid holding 100\% across the
entire range (zero denied entries).

\begin{figure}[!t]
\centering
\includegraphics[width=0.85\columnwidth]{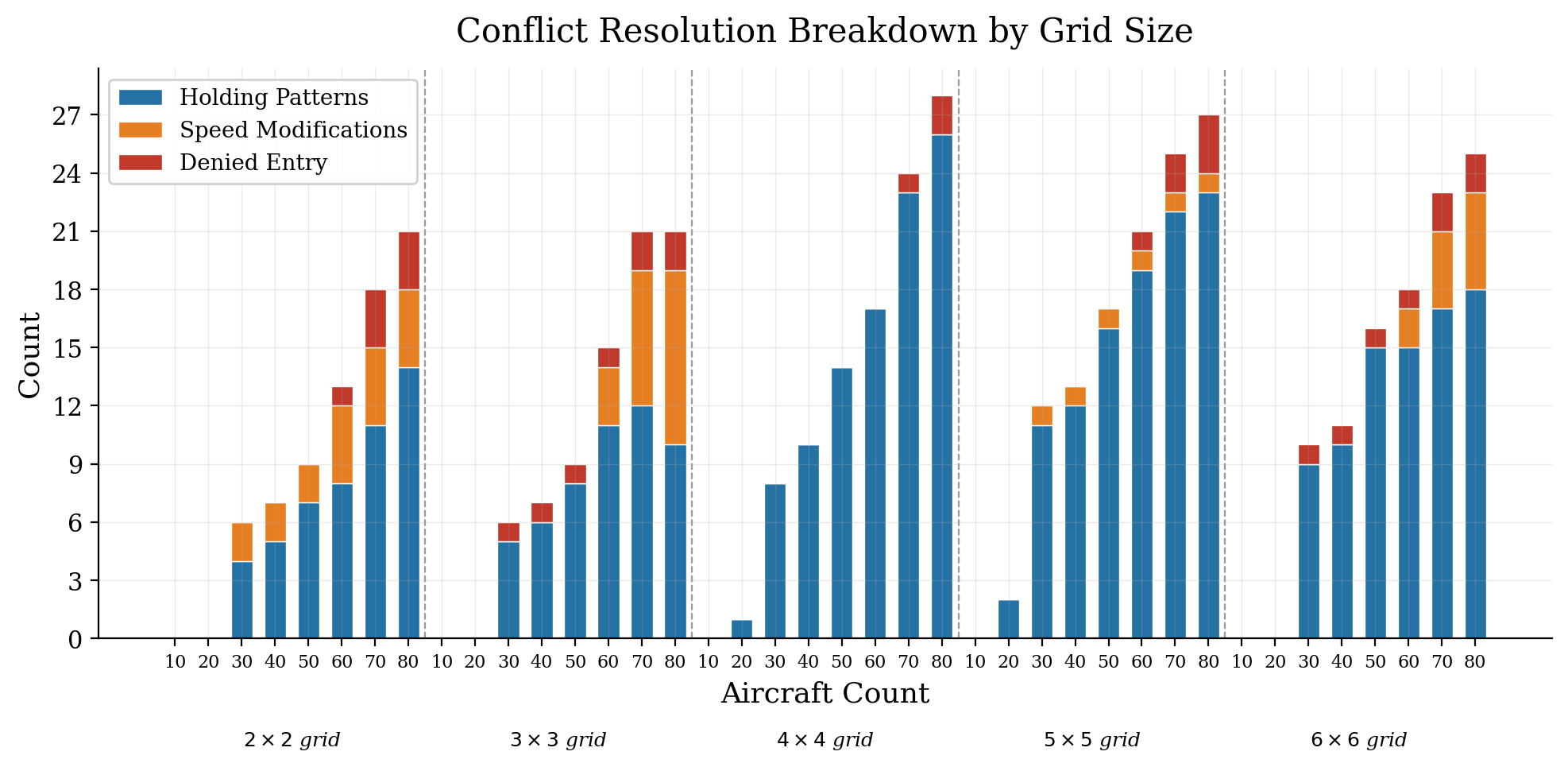}
\caption{Conflict resolution breakdown for JFK real world traffic.
Stacked bars show holdings, speed modifications, and denied entries.
}\label{fig:conflict}
\end{figure}

Fig.~\ref{fig:conflict} shows the conflict resolution breakdown for
the JFK sweep.  Holding patterns are the most frequently used
mechanism because speed modification, though attempted first, often
fails to resolve geometric conflicts.  Denied entries remain a small
fraction ($\leq 3$ across all JFK configurations; the synthetic sweep
shows a similarly small fraction of $\leq 12$),
confirming that the holding plus speed mechanism resolves the vast
majority of conflicts.

\begin{figure}[!t]
\centering
\includegraphics[width=0.85\columnwidth]{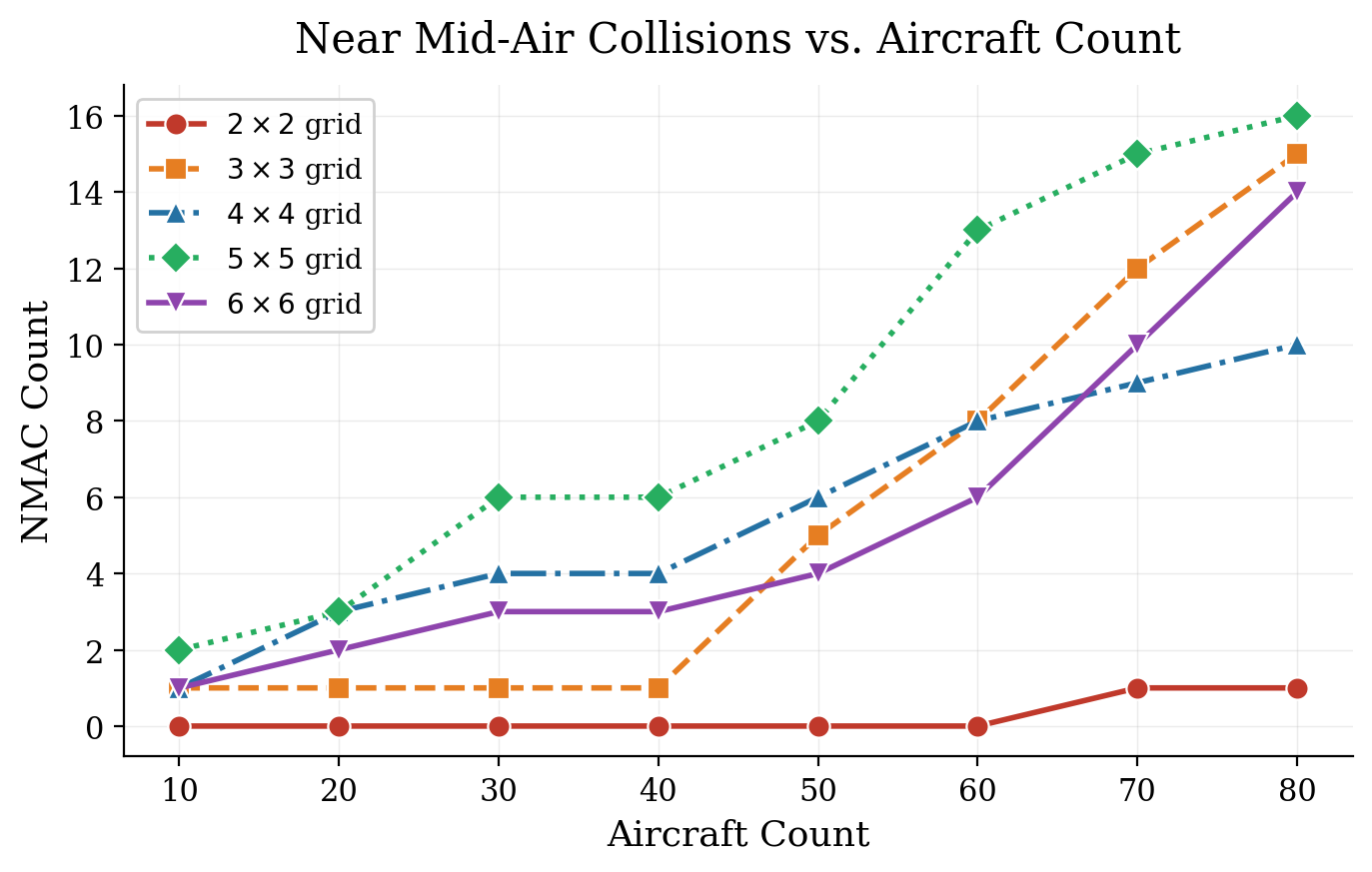}
\caption{NMACs vs.\ aircraft count for JFK real world traffic.  The
$2{\times}2$ grid produces the fewest NMACs because most converging
approach traffic stays within a single sector.  }\label{fig:nmac}
\end{figure}

Fig.~\ref{fig:nmac} shows Near Mid-Air Collision (NMAC) counts for
the JFK sweep: the
$2{\times}2$ grid produces the fewest NMACs (0--1 across all counts)
because most converging approach traffic stays within a single sector
and is handled by one consensus round, while finer grids split
approach corridors across sector boundaries, creating inter sector
separation violations.  This relationship reverses in the synthetic
sweep, where coarser grids
produce the most NMACs (40 at 160 aircraft for $2{\times}2$, down to 8
for $16{\times}16$) because fewer internal boundaries means fewer
opportunities for the protocol to detect converging traffic.  The
reversal highlights that the optimal grid size for safety depends on
traffic geometry, not just density, motivating the use of the XGBoost
predictor to adapt the grid to actual traffic patterns.

\begin{figure}[!t]
\centering
\includegraphics[width=0.85\columnwidth]{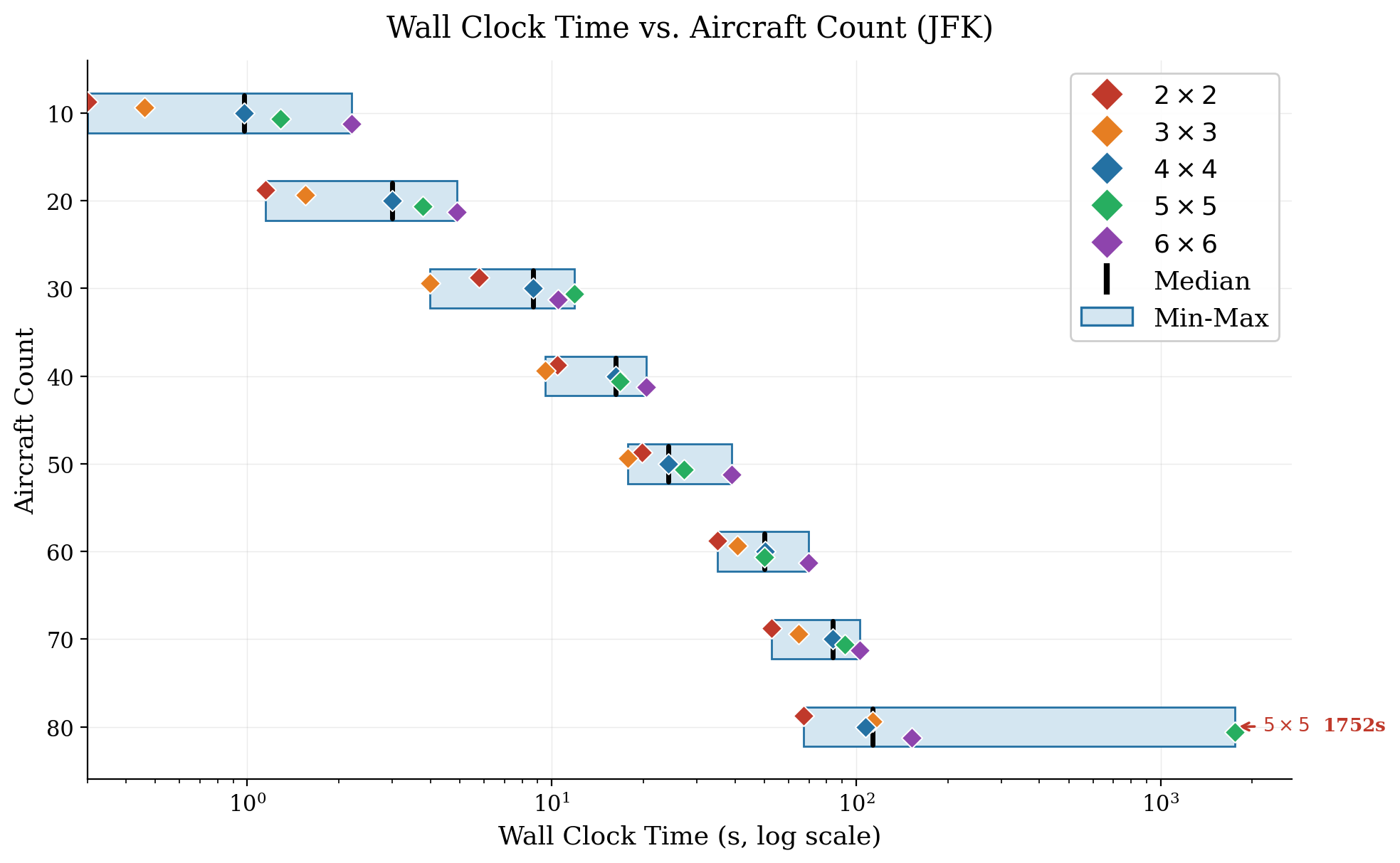}
\caption{Wall clock time vs.\ aircraft count (JFK).  Horizontal bars
show min to max range across grid sizes.  The $5{\times}5$ outlier at 80
aircraft (1,752\,s) is caused by Paxos livelock in a single
sector.}\label{fig:wallclock}
\end{figure}

Fig.~\ref{fig:wallclock} shows wall clock execution time for the JFK
sweep as a Gantt style range chart.  Each horizontal bar spans the min
to max time across grid sizes at a given aircraft count, with diamond
markers for individual grids and a vertical line at the median.  Times
scale super linearly with aircraft count, driven by quadratic growth in
pairwise conflict checks.  Most configurations at 80 aircraft complete
in under 200\,s.  The striking outlier is the $5{\times}5$ grid at 80
aircraft (${\sim}1{,}752$\,s), caused by Paxos livelock in a single
high density sector.  This occurred once across all experiments,
confirming it as a tail risk rather than a systemic limitation.

\begin{figure}[!t]
\centering
\includegraphics[width=0.85\columnwidth]{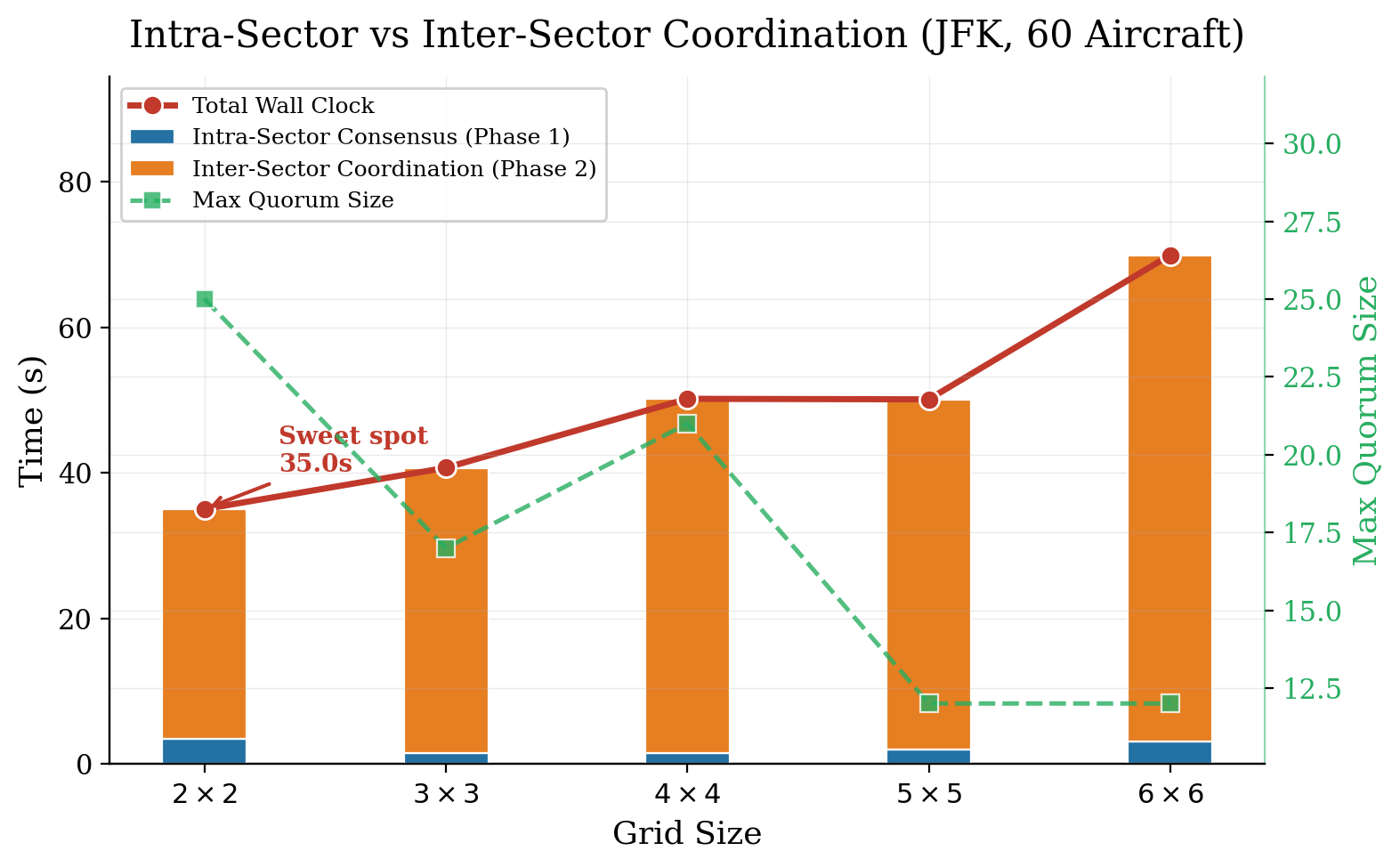}
\caption{Intra-sector consensus (Phase~1) vs.\ inter-sector
coordination (Phase~2) time as a function of grid size (JFK,
60~aircraft).  Phase~2 dominates and grows with finer grids; the
secondary axis shows the corresponding decrease in maximum quorum
size.}\label{fig:tradeoff}
\end{figure}

Fig.~\ref{fig:tradeoff} decomposes execution time into its two phases
to expose the fundamental sectorization tradeoff.  Phase~1
(intra-sector consensus) remains under 4\,s across all grid sizes
because finer grids reduce quorum sizes, making each Paxos round
faster.  Phase~2 (inter-sector coordination) dominates and grows
monotonically: more sectors mean more boundary crossings and handoff
negotiations, rising from 31.6\,s at $2{\times}2$ to 66.7\,s at
$6{\times}6$.  Total wall clock time is therefore minimized at the
coarsest grid ($2{\times}2$, 35.0\,s), but as shown in
Fig.~\ref{fig:nmac} this configuration incurs the highest NMAC rate.
The $4{\times}4$ grid represents the practical sweet spot: it keeps
total time within 43\% of the minimum while eliminating all NMACs at
60~aircraft.

\section{Bayesian Optimization}

\subsection{Formulation}

The consensus protocol exposes eight tunable parameters: timeout
duration, number of retry attempts, number of alternates, phase delay
factor, NACK backoff minimum and maximum, Initial Request (IR) time,
and solver timeout.  Even a coarse five level discretization of each
parameter would require $5^{8} = 390{,}625$ simulations.  Each
evaluation requires a full 30 minute simulation, making gradient free
black box optimization the appropriate
regime~\cite{shahriari_bo_survey_2016}.

The objective combines four terms:
\begin{equation}\label{eq:score}
J(\theta) = 100 \cdot r_{\text{success}} - 30 \cdot r_{\text{hold}} - 10 \cdot r_{\text{speed}} - 2 \cdot \bar{n}_{\text{retry}}
\end{equation}
where $r_{\text{success}}$ is the admission throughput ratio,
$r_{\text{hold}}$ and $r_{\text{speed}}$ are the fractions of admissions
requiring holding patterns or speed modifications, and
$\bar{n}_{\text{retry}}$ is the average retry count per initiated
request.  The coefficients were chosen based on operational reasoning:
entry success is the dominant objective ($\times 100$), holding consumes
more fuel than speed changes ($\times 30$ vs.\ $\times 10$), and retries
are invisible to end users ($\times 2$).  Hard penalties of $-100$ for
any NMAC and $-50$ for wall clock timeout override the continuous score.
Derivations of the weight choices and per-airport sensitivity studies
appear in the thesis.

A Gaussian Process (GP) with a Mat\'{e}rn~5/2 Automatic Relevance
Determination (ARD) kernel~\cite{rasmussen_gp_2006} models the
objective surface.  The ARD
parameterization assigns a separate lengthscale $\ell_d$ to each input
dimension; after fitting, dimensions with shorter lengthscales exhibit
greater influence.  Expected
Improvement~\cite{jones_ego_1998, snoek_practical_bo_2012} selects each
trial.  The budget is 50 trials per airport: 10 Latin Hypercube
Sampling (LHS)~\cite{mckay_lhs_1979} initial samples followed by 40 GP
guided iterations.

An initial campaign on ORD produced five timed out trials at parameter
extremes.  We tightened bounds to exclude the pathological region and
enforce \texttt{nackBackoffMin} $<$ \texttt{nackBackoffMax}.

\subsection{Results}

Table~\ref{tab:bo} summarizes optimal configurations across three
airports with different traffic densities.

\begin{table}[!htb]
\centering
\caption{Best BO configurations by airport.}\label{tab:bo}
\renewcommand{\arraystretch}{1.2}
\begin{tabular}{lrrr}
\toprule
 & \textbf{LAX} & \textbf{ORD} & \textbf{DFW} \\
 & (31 acft) & (47 acft) & (60 acft) \\
\midrule
Best Score        & 189.8 & 127.1 & 100.8 \\
Timeout Duration (s) & 0.50  & 0.50  & 2.93  \\
IR Attempts       & 10    & 15    & 4     \\
Solve Timeout (s) & 10.0  & 1.0   & 9.1   \\
Timeouts (/50)    & 0     & 5     & 1     \\
\bottomrule
\end{tabular}
\end{table}

\begin{figure}[!htb]
\centering
\includegraphics[width=0.85\columnwidth]{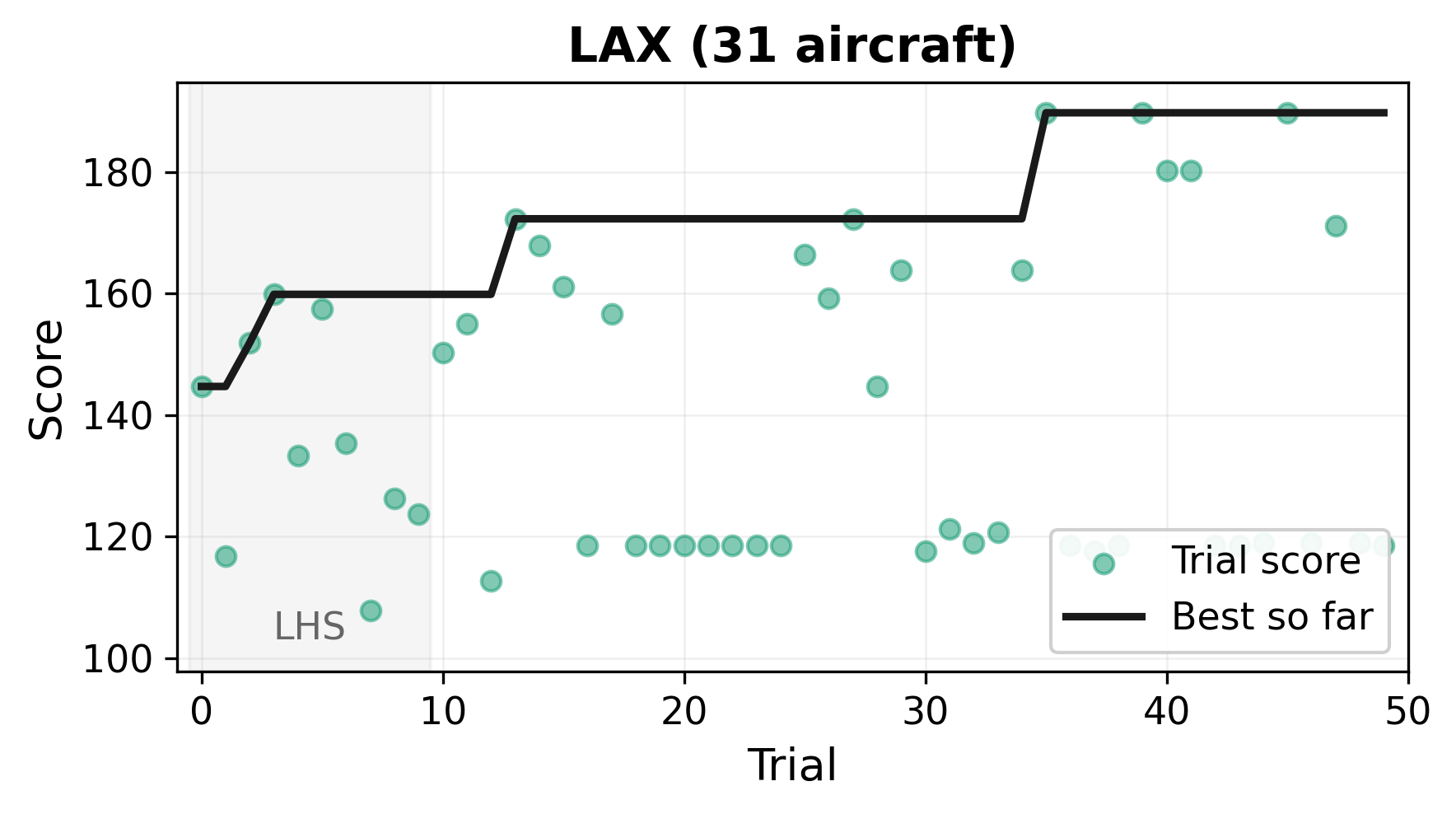}
\caption{BO convergence trajectory for LAX (31 aircraft).}\label{fig:bo-lax}
\end{figure}

\begin{figure}[!htb]
\centering
\includegraphics[width=0.85\columnwidth]{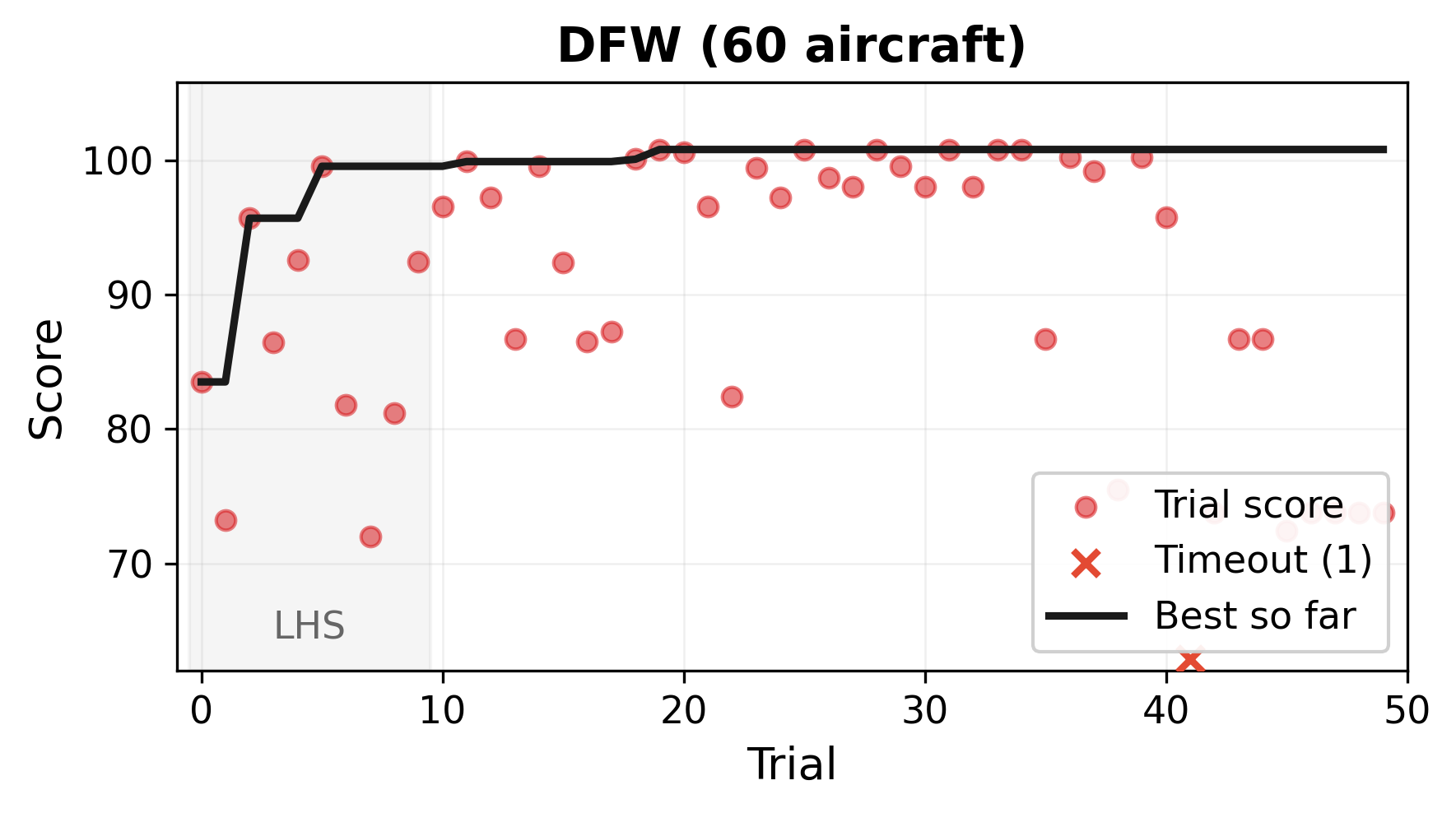}
\caption{BO convergence trajectory for DFW (60 aircraft).  }\label{fig:bo-dfw}
\end{figure}

Figs.~\ref{fig:bo-lax} and~\ref{fig:bo-dfw} show convergence
trajectories for LAX and DFW.
Across all three airports, LAX reaches its best score at trial~35, ORD
at trial~40, and DFW at trial~19.  The optimal configurations differ
qualitatively: LAX favors many retry attempts with a long solver
window, ORD requires a short solver timeout (1.0\,s) to avoid
pathological retry chains that caused 5 of its 50 trials to time out,
and DFW favors few retries (4) with a longer timeout duration
(2.93\,s).  This confirms that no single default setting works and
that protocol tuning must be traffic density dependent.

\begin{table}[!htb]
\centering
\caption{Top three parameter importances per airport, derived from GP
ARD lengthscales (normalized inverse lengthscale).}\label{tab:importance}
\renewcommand{\arraystretch}{1.2}
\footnotesize
\begin{tabular}{llr}
\toprule
\textbf{Airport} & \textbf{Parameter} & \textbf{Importance} \\
\midrule
\multirow{3}{*}{LAX (31 acft)}
 & Solve Timeout     & 36.7\% \\
 & NACK Backoff Min  & 31.7\% \\
 & Start IR Time     & 11.4\% \\
\midrule
\multirow{3}{*}{ORD (47 acft)}
 & NACK Backoff Min  & 64.3\% \\
 & Start IR Time     & 27.6\% \\
 & Timeout Duration  & \phantom{0}3.1\% \\
\midrule
\multirow{3}{*}{DFW (60 acft)}
 & Timeout Duration  & 45.6\% \\
 & NACK Backoff Min  & 23.2\% \\
 & Alternates        & 14.8\% \\
\bottomrule
\end{tabular}
\end{table}

Table~\ref{tab:importance} summarizes the top three parameter
importances per airport, derived from the ARD kernel's per parameter
lengthscales.  Solver timeout dominates at LAX (36.7\%), while NACK
backoff and timeout duration become more important at higher densities
where contention drives retry behavior.

\section{Discussion}

Three specific findings stand out.

First, the XGBoost predictor achieves 91.38\% accuracy using only
aggregate traffic features with no geographic identifiers, confirming
that sectorization depends on traffic state rather than airspace
identity and supporting deployment to unseen regions without retraining.
Predicting from aggregate features also scales to larger configuration
spaces where exhaustive scoring becomes impractical.

Second, the NMAC analysis reveals that grid size interacts with traffic
geometry in non-obvious ways.  Uniform grids that work well for random
traffic can perform poorly on real approach corridors, and vice versa.
This motivates the use of the predictor to select grid configurations
adapted to actual traffic patterns rather than relying on fixed
partitions.

Third, the livelock outlier at $5{\times}5$/80 aircraft underscores why
automatic parameter tuning is essential.  BO identified NACK backoff as
the most important parameter at ORD and DFW (64.3\% and 23.2\%
importance), precisely the mechanism that prevents livelock.  Manual
tuning is unlikely to find these configurations given the
eight-dimensional search space.

Furthermore, Bayesian Optimization confirms that protocol sensitivity is
environment dependent.  The qualitative differences between optimal
configurations at LAX, ORD, and DFW mean that manual parameter selection
cannot scale.  Automatic tuning is essential for deployment across
diverse environments.

\textbf{Limitations.}
The simulation uses straight line flight paths and point mass dynamics,
which limits realism near terminal areas where aircraft follow curved,
altitude constrained procedures.  However, the consensus protocol is
trajectory agnostic: it operates on the sequence of waypoints and
velocity vectors an aircraft submits, not on how those waypoints were
produced.  Replacing the straight line generator with a procedural or
published flight track therefore requires no protocol changes.  The
same admission, conflict detection, and TAP dissemination logic
applies to any flight plan representation.  Communication is
deterministic, reliable, has no faulty processes, and no latency jitter.
The protocol does not currently handle aircraft emergencies, which would
require priority override mechanisms.  The BO budget of
50 trials per airport may miss fine grained optima.  The objective
function weights (100, 30, 10, 2) were set by intuition rather than
derived from data.

\section{Future Work and Conclusion}

\textbf{Future work.}
Submillisecond inference makes periodic re-sectorization feasible every
few minutes as traffic patterns shift, with hysteresis to prevent
oscillation.

Testing under realistic communication models with packet loss and
ADS-B latency distributions would quantify the gap between ideal and
deployed performance, particularly the effect of packet loss on retry
rates.

Learning the BO objective weights from data via bilevel optimization or
inverse reinforcement learning would remove the need for manual
coefficient selection and could reveal that the relative importance of
success rate, holding, speed modifications, and retries differs from
what intuition suggests.

\textbf{Conclusion.}
We presented a three stage pipeline for adaptive airspace sectorization
and decentralized admission control.  A two stage XGBoost predictor
selects optimal grid configurations at 91.38\% accuracy from 23 location
agnostic features.  A Paxos consensus protocol coordinates sector entries
with above 96\% success and low NMAC rates across both sweeps.
Bayesian Optimization automatically tunes protocol parameters, confirming
that each airport requires its own configuration.  The submillisecond
prediction time and location-agnostic design make the system a practical
candidate for autonomous airspace management; \textbf{codebase:}~\cite{dhodapkar_datc_code_2025}.

\bibliographystyle{IEEEtran}
\bibliography{smc_references}

\end{document}